\documentclass[12pt,amsfonts]{revtex4}
\usepackage{amsmath}
\usepackage{graphicx}
\usepackage{psfrag}
\renewcommand{\slash}{\! \not \!}
\begin{document}
\title{One loop effective potential in heterotic $M$-theory}
\author{Ian G. Moss}
\email{ian.moss@ncl.ac.uk}
\author{James P. Norman}
\email{j.p.norman@ncl.ac.uk}
\affiliation{School of Mathematics and Statistics, University of Newcastle upon Tyne,
Newcastle upon Tyne, NE1 7RU, United Kingdom}
\date{\today}
\begin{abstract}
We have calculated the one loop effective potential of the vector multiplets
arising from the compactification to five dimensions of heterotic $M$-theory 
on a Calabi-Yau manifold with $h^{1,1}>1$. We find that extensive cancellations
between the fermionic and bosonic sectors of the theory cause the effective
potential to vanish, with the exception of a higher order curvature term of the
type which might arise from string corrections.
\end{abstract}
\maketitle

\section{Introduction}

The weakly coupled $E_8 \times E_8$ heterotic string is one of the most 
phenomenologically viable of the five consistent string theories. 
Unfortunately, the predicted value for Newton's constant
in this theory is too large. Witten \cite{Witten:1996mz} has shown that this
situation can be resolved in the
\emph{strong} coupling limit, which is believed to be eleven-dimensional
supergravity on the
orbifold ${\cal{M}}^{10} \times S^{1}/Z_2$, with $E_8$ Super Yang-Mills
gauge theories on the orbifold fixed points \cite{Horava:1996qa}. This
theory can be compactified on a Calabi-Yau three-fold to obtain six internal
dimensions.
It is known that in order for the theory to predict the correct known values of
Newton's constant
and grand unification gauge couplings, the orbifold radius must be an order
of magnitude or so larger than the Calabi-Yau compactification scale. Hence,
at some intermediate energy scale, the theory has a consistent five-dimensional
description.

Lukas {\it et al.} \cite{Lukas:1998yy, Lukas:1998tt} have derived the
five-dimensional effective
action by reducing Ho\v rava-Witten theory on a Calabi-Yau space.
They have shown that the resulting theory 
is a gauged version of $N=1$ supergravity in five dimensions, with a
non-abelian set of $E_8$ gauge fields on one orbifold plane, spontaneously
broken
to $E_6$, and an unbroken 
$E_8$ gauge group on the other. The vacuum solution for this theory has a
domain wall structure with a curved bulk metric. This was the predecessor of 
the ``brane-world'' scenarios \cite{Randall:1999ee}. 

In addition to a gravitational and universal hypermultiplet present for
every Calabi-Yau compactification, 
the five-dimensional theory contains a number of vector multiplets, depending
on the topological properties of the Calabi-Yau. More precisely, there
are $n_V=h^{1,1}-1$ of these vector multiplets. The background solution is
a multi-charged BPS domain wall with $h^{1,1}$ charges. Lukas {\it et al.} show
that
there always exists a
background solution with $h^{1,1}\ge1$ with two moduli parameters representing
the separation of the domain walls and the dilaton field.

In this paper, we will calculate the one loop effective potential of these
vector multiplets about this simple background solution as a function of the
moduli.

The vector multiplets decouple from the gravitational and universal
hypermultiplets
making them the simplest multiplets for the one loop calculation. Furthermore,
it is possible to
change the number of vector hypermultiplets by changing $h^{1,1}$ without
altering the
background solution or any of the other multiplets at one loop order.
Therefore, the
vector multiplets tell us how the total vacuum energy depends on the topology
of the
Calabi-Yau manifold.

The five dimensional domain wall solution gives rise to an effective four
dimensional
theory in which the separation of the domain walls becomes one of the moduli
fields. It
is important to identify effects which can provide a potential for the brane
separation
and fix this particular modulus. One possible mechanism is that quantum
fluctuations of the bulk fields stabilise the branes at
phenomenologically acceptable positions. This has been discussed extensively
in the context of the Randall-Sundrum brane world scenario 
\cite{Goldberger:2000dv,Garriga:2000jb,Flachi:2001pq,Flachi:2001ke,Flachi:2001bj}.
Previous work of this kind in five-dimensional heterotic $M$-theory
has been done for scalar fields by Garriga et al. \cite{Garriga:2001ar}. 

The BPS solution described above preserves half of the
supersymmetry, hence one would expect cancellation of the contributions
to the effective action of the bosonic and fermionic sectors. 
In a related calculation in an eleven
dimensional setting, this expectation was confirmed when the vacuum energy was
shown to be zero \cite{Fabinger:2000jd}. However, the Ho\v rava-Witten theory
has only been constructed so far as an expansion
in powers of the eleven dimensional gravitational coupling constant $\kappa$,
and it is
not known whether there exists a fully supersymmetric formulation valid to all
orders
in $\kappa$.

We evaluate the
one-loop effective action of the various fields using $\zeta$-function 
regularisation. We have extended the conformal transformation technique
\cite{Dowker:1990ue} used in \cite{Garriga:2001ar} to apply to spin $1/2$ and
spin $1$ fields. 

In the following, rather than working in the ``upstairs'' picture of the
orbifold $S^1/Z_2$, we work in the equivalent ``downstairs'' description
of an interval $I$. That is, we consider the space to be a manifold with a
boundary, rather than a manifold with singular delta-function sources. In this
way, the domain walls in the above are replaced by boundary ``branes''.

Our conventions are as follows: Tensor indices are 
$\alpha$,$\beta$,$\gamma$, $\ldots = 0,1,2,3,11$ in the bulk, and 
$\mu$,$\nu$,$\rho$, $\ldots = 0,1,2,3$ on the boundary. The
coordinate in the direction of the interval is $x^{11}$. Capital letters
$A$,$B$,$C$,$\ldots = 0,1,2,3,11$ index a local orthonormal frame in the
bulk, while $I$,$J$,$K$,$\ldots = 0,1,2,3$ label a local orthonormal frame
on the boundary. 
Conventions for the Riemann and Ricci tensors are as in Misner, Thorne and
Wheeler \cite{MTW}.
The extrinsic curvature is defined as $K_{\mu \nu}=\nabla_\mu N_\nu$ where
$N_\alpha$ is the unit \emph{outward} pointing normal. Gamma matrices are
defined by $\left\{\Gamma^\alpha,\Gamma^\beta\right\}=2g^{\alpha \beta}$. 
The index
$N$ denotes contraction with $N_\alpha$. 

\section{Classical background solution}\label{sec:classical}

The five dimensional action of heterotic $M$-theory including the $(1,1)$
moduli was derived in \cite{Lukas:1998tt}. Here we will briefly review the
field content and background solution.

The background solution consists of setting as many of the fields as possible
to zero. The minimum field content consistent for 
describing the vacuum solution is given by the action
$S^{0}=S^{0}({\cal M})+S^{0}(\partial {\cal M})$,
where the bulk and boundary actions are
\begin{eqnarray}
S^{0}({\cal M})&=&-\frac{1}{\kappa_5^2}\int_{\cal M} d\mu \left(-R
+G_{ij}\partial_\alpha b^i
\partial^\alpha b^j +\frac12 \partial_\alpha \Phi \partial^\alpha \Phi
+\frac12 e^{-2 \Phi} G^{ij} \alpha_i \alpha_j \right).\\
S^{0}(\partial {\cal M})&=&\frac{2}{\kappa_5^2}\int_{\partial{\cal M}^{(1)}} 
d \mu\left(K+\frac{\sqrt{2}}{2} \alpha_i b^i e^{-\Phi}\right)
                         +\frac{2}{\kappa_5^2}\int_{\partial{\cal M}^{(2)}} 
d \mu\left(K-\frac{\sqrt{2}}{2} \alpha_i b^i e^{-\Phi}\right).
\end{eqnarray}
Let us first remind the reader of the origin of some of these fields. The
scalar field $\Phi$ is the moduli field describing the volume of Calabi-Yau and
will be referred to as the dilaton. 
The fields $b^i$ are the Calabi-Yau shape moduli. Indices $i$,$j$,$\ldots$ run
from $1,\ldots, h^{1,1}$ and are raised
and lowered by the Kahler metric $G_{ij}=-\frac12 \partial_i \partial_j \ln
{\cal K}$, where ${\cal K}=d_{ijk}b^ib^jb^k$ is the Kahler potential. The
constants $d_{ijk}$ are the Calabi-Yau intersection numbers.
The charges $\alpha_i$ are constants. The shape moduli are constrained by
${\cal K}=6$ (after the differentiation)
and hence only represent $h^{1,1}-1$ degrees of freedom.

The existence of the bulk and boundary 
potentials leads to a curved bulk metric and to non-trivial profiles of $\Phi$ 
and $b^i$ across the orbifold. We denote $y=x^{11}$. The position of the
visible brane is taken to be at $y=0$ and the hidden brane is at $y=\pi$.
Upon substitution of the ansatz
\begin{eqnarray}
ds^2&=&a(y)^2\eta_{\mu \nu}d x^\mu d x^\nu + b(y)^2 dy^2 \\
\Phi&=&\Phi(y)\\
b^i&=&b^i(y),
\end{eqnarray}
Lukas {\emph et al.} have found an implicit solution to the equations of motion 
in terms of $h^{(1,1)}$ functions $f^i(y)$. However, they also show that by 
defining new constants
$\bar{\alpha}^i$ and $\alpha$ by
\begin{equation}
d_{ijk}\bar{\alpha}^i\bar{\alpha}^j=\frac23 \alpha^i, \quad
\alpha= 9 \left(\frac16 d_{ijk}
\bar{\alpha}^i\bar{\alpha}^j\bar{\alpha}^k\right),
\end{equation}
and by using an appropriate choice of integration constants, one can find the 
explicit solution,
\begin{eqnarray}
a & = &a_0 H^{1/2} \\
b & = & b_0 H^{2} \\
\Phi & = & \ln \left( b_0 H^3\right) \\
b^i&=& \frac{3}{2}\frac{\alpha^i}{\alpha}
\end{eqnarray}
where
\begin{equation}
H=\frac{\sqrt2}{3} \alpha y + c_0,
\end{equation}
which is just the ``universal'' solution obtained when $h^{1,1}=1$. 
In this classical solution, the fields $b^i$ are constant across the orbifold.

\section{One-loop effective action}

To evaluate the one-loop correction to the classical action for a field
$\phi_a$, we perform a continuation of the Lorentzian metric to a positive 
definite Euclidean metric. We expand the Euclidean action $I$ around the 
background fields $\phi^a_0$, i.e.,
$I=I(\phi_0)+I^{,a}(\phi_0)(\phi-{\phi_0})_a
+I^{,a}_{\ \ \ b}(\phi_0) (\phi-\phi_0)_a(\phi-\phi_0)^b+\ldots$.
The one-loop correction to the classical action is
\begin{equation}
W=(-1)^f\frac12 \log \det \left( \mu_R^{-2} \Delta^{a}_{\ \ b} \right)
\end{equation}
where we use comma ($,$) to denote functional differentiation with respect to
the field $\phi_a$. $f=0$ for bosons and $f=1$ for fermions.
$\Delta^{a}_{\ \ b} = I^{,a}_{\ \ b}$ is an operator. 
We define the determinant of
an operator using generalised
$\zeta$-functions. The $\zeta$-function is defined by a generalised trace,
\begin{equation}
\zeta(s)={\rm tr}\left(\Delta^{-s}\right).
\end{equation}
for some range of $s$ in the complex plane where the trace converges. We
define the determinant of the operator by
\begin{equation}
\log \det \left(\mu_R^{-2} \Delta^a_{\ b}\right)=-\zeta'(0)-\zeta(0)\log \mu_R^{2}
\end{equation}
where we have used the analytic continuation of the $\zeta$-function at $s=0$.

In curved space, the trace of an operator is often difficult to evaluate.
However, we can use the properties of the effective action under conformal
transformations of the metric 
\begin{equation}\label{eq:contrans}
g_{\mu \nu}\rightarrow \tilde{g}_{\mu \nu} = \Omega^2 g_{\mu \nu},
\end{equation} 
to
relate the effective actions of an operator in conformally related spacetimes.
We use a tilde to denote quantities calculated in the metric $\tilde{g}$.
We restrict attention here to second order operators of Laplace type  -- that
is, operators which can be written in the form
\begin{equation}\label{eq:laplace}
\Delta=-D^2+X,
\end{equation}
where $D_\alpha=\nabla_\alpha+iA_\alpha$ is a covariant gauge derivative.
We use the $n$-bein formalism to treat fields of general spin. 
The covariant derivative is 
$\nabla_\alpha=\partial_\alpha+i\omega_\alpha^{\ AB}\Sigma_{AB}$
where $\Sigma_{AB}$ is the generator of Lorentz transformations for the
appropriate
representation of the Lorentz group. Under a conformal
rescaling (\ref{eq:contrans}), the $n$-bein and its dual rescale as
\begin{equation}
\tilde{e}^A_\alpha = \Omega e^A_\alpha, \quad
\tilde{e}_A^\alpha = \Omega^{-1} e_A^\alpha,
\end{equation}  
while the operator (\ref{eq:laplace}) rescales to $\tilde{\Delta}$, where  
\begin{equation}
\tilde{\Delta} = -\left(\tilde{\nabla}
+i\tilde{A}\right)^2+
\tilde{X}.
\end{equation} 
If we define the conformal rescaling properties of $\tilde{X}$ and
$\tilde{A}$ as
\begin{equation}\label{eq:xtrans}
\tilde{X} = \Omega^{-2} X -\frac{3}{16}\left(\Omega^{-2} R - 
\tilde{R} \right)
,\quad \tilde{A}_{\alpha}=A_\alpha-2\Omega^{-1}\Omega_{;\beta}
\Sigma_\alpha^{\ \beta},
\end{equation}
then the rescaled operator $\tilde{\Delta}$ is given by
\begin{equation}
\tilde{\Delta} = \Omega^{-7/2} \Delta \Omega^{3/2}.
\end{equation}

Boundary conditions will also be affected by conformal transformations. If
the field satisfies Robin boundary conditions
\begin{equation}
\left( D_N -\cal{S} \right) \phi =0 \quad \textrm{on} \quad \partial {\cal M},
\end{equation}
then under a conformal rescaling, this will become
\begin{equation}
\left( \tilde{D}_N - \tilde{\cal S} \right) \tilde{\phi}=0 \quad \text{on} 
\quad \partial {\cal M}.
\end{equation} 
since the field transforms as $\phi\rightarrow \tilde{\phi} = \Omega^{-3/2}
\phi$. The boundary conditions retain the same form if we define
\begin{equation}\label{eq:strans}
\tilde{\cal{S}}=\Omega^{-1} {\cal{S}} + \frac38 
\left(\Omega^{-1}K-\tilde{K}\right).
\end{equation}
Dirichlet boundary conditions $\phi=0$ on $\partial {\cal M}$ are unchanged
under conformal rescalings.

Following Dowker \cite{Dowker:1990ue}, we introduce a one parameter family of
metrics which
interpolate between two conformally related spacetimes. 
We take $\Omega=\Omega(\sigma)$.
One can then show that, for operators which transform covariantly under
conformal transformations,
\begin{equation}
W\left[\sigma=\sigma_2\right]=W\left[\sigma=\sigma_1\right]
+C\left[\Omega \right],
\end{equation}
where the cocycle function $C\left[\Omega \right]$ is given (in five
dimensions) 
in terms of
the generalised heat kernel coefficient $a_{5/2}(p,\Delta)$ as
\begin{equation}
C[\Omega]=-(-1)^f\int_{\sigma_1}^{\sigma_2} d\sigma \ 
a_{5/2}\left(\partial_\sigma \ln
\Omega(\sigma),\tilde{\Delta}\right).\label{cocycle}
\end{equation}
The $a_{5/2}\left(p,\Delta\right)$ coefficient is 
known for general Laplace type
operators with mixed boundary conditions \cite{Branson:1999jz}. 
It is composed of geometric invariants in the metric
evaluated only on the boundary of the spacetime. Hence, we can relate the
effective actions of the conformally related operators in two conformally
related spacetimes.

The classical background in Section~\ref{sec:classical} is conformally flat
in the coordinate $z$ defined by $dz=a(y)/b(y)dy$. In this new coordinate,
the metric is
\begin{equation}
ds^2 = \left(\frac{z}{z_1}\right)^{2/5} \left(\eta_{\mu \nu}dx^\mu
dx^\nu+dz^2\right),
\end{equation}
where
\begin{equation}\label{eq:z1}
z_1={3\sqrt{2} b_0\over 5\alpha a_0}.
\end{equation}
The dilaton field,
\begin{equation}
\Phi=\frac65\log\left(\frac{z}{z_1}\right)+
\log\left(\frac{5\alpha z_1}{3\sqrt{2}}\right).
\end{equation}
The values of $z$ on the two branes, $z_1$ and $z_2$ can be used as the moduli
parameters of the background solution. We can use the technique described
above to relate the one-loop effective action in the warped background
spacetime to one in flat space. 

For definiteness, we take the function $\Omega(\sigma)$
to be
\begin{equation}
\Omega(\sigma) = e^{(1-\sigma) \omega(z)}
\end{equation}
so that flat space is at $\sigma=0$ and the physical metric is at $\sigma=1$.
Then, the effective action is related to the flat space effective action $W_0$
by
\begin{equation}
W=W_0+C\left[\Omega\right], \quad 
C[\Omega]=(-1)^f\int_0^1 d\sigma \ 
a_{5/2}\left(\omega,\tilde{\Delta}\right).
\end{equation}
From now on we use the subscript ``0'' to indicate a quantity calculated in
the conformally transformed flat space.

\section{The one loop effective action of the vector multiplet}

\subsection{The scalar field}\label{sec:scalar}

There are $h^{1,1}$ scalar fields $b^i$ in the vector multiplet. However, these
are constrained, and as such only represent $n_V=h^{1,1}-1$ degrees of freedom.
We can, however, write the action in terms of $n_V$ unconstrained fields
$\varphi_x$. The indices $x$, $y$, etc are raised and lowered using
the metric $g_{xy}=b^i_x b^j_y G_{ij}$. The action contains both bulk and
boundary terms 
$S=S({\cal M})+S(\partial{\cal M})$, which are
\footnote{From now on we set $k_5^2=1$}
\begin{eqnarray}
S({\cal M})&=-&\int_{\cal M} d\mu \left( \frac12 g_{xy} \partial_\alpha
\varphi^x \partial^\alpha \varphi^y + V(\varphi) \right)\\
S(\partial {\cal M})&=&\int_{\partial {\cal M}^{(1)}} d\mu \ U(\varphi)-
\int_{\partial {\cal M}^{(2)}} d\mu \ U(\varphi).
\end{eqnarray}
The bulk potential $V(\varphi)$ and the boundary potential $U(\varphi)$ are 
only
expressible implicitly in terms of the constrained fields $b^i$. They are
\begin{equation}
V(\varphi) = \frac14 e^{-2\Phi} G^{ij} \alpha_i \alpha_j, 
\quad U(\varphi)= \frac{\sqrt{2}}{2} e^{-\Phi} \alpha_i b^i. 
\end{equation}
The second variation of the classical action about the background solution
gives the operator
\begin{equation}\label{eq:scalarop}
\Delta^{x}_{\ y}= -\nabla^2 \delta^{x}_{\ y} + V(\varphi)^{,x}_{\ y}
\end{equation}
together with the boundary conditions
\begin{equation}\label{eq:scalarbcs}
\left(\nabla_N \delta^{x}_{\ y} - U(\varphi)^{,x}_{\ y}
\right) \varphi^y =0 \quad \textrm{on} \quad \partial {\cal M}^{(1)}, \quad
\left(\nabla_N \delta^{x}_{\ y} + U(\varphi)^{,x}_{\ y}
\right) \varphi^y =0 \quad \textrm{on} \quad \partial {\cal M}^{(2)}.
\end{equation}
Although the bulk and boundary potentials are only known in terms of the
constrained fields $b^i$, the second derivatives of the bulk and boundary 
potentials with respect to the fields $\varphi^x$ can be calculated in terms
of the background scalar field $\Phi$.
\begin{equation}
V(\varphi)_{,xy} = \frac49 \alpha^2 e^{-2\Phi} g_{xy}, 
\quad U(\varphi)_{,xy}= \frac{\sqrt{2}}{3}\alpha e^{-\Phi} g_{xy}
\end{equation}
Since the operator has a trivial index structure in the indices $x,y$, from
this point we drop this index to avoid clutter. 
The operator (\ref{eq:scalarop}) is of Laplace type (\ref{eq:laplace}). 
We now perform a conformal transformation (\ref{eq:contrans}) with 
$\Omega=e^{\omega(z)}$.  The one-loop effective action in the conformally
transformed space requires the eigenvalues of the conformally transformed
operator $\Delta_0 \tilde\varphi_n = \lambda_n \tilde\varphi_n$.
The general solution can be found in terms of Bessel functions;
\begin{equation}
\tilde{\varphi}_n(z)=\sqrt{z}e^{ik_\mu x^\mu}\left(A J_{3/5}(m_n z)+B
Y_{3/5}(m_n z) \right),
\end{equation}
where $\lambda_n= m_n^2+k^2$.
We take the position of the visible brane to be at $z_1$ and the hidden
brane to be at $z_2$ ($z_2>z_1$). We also introduce $\tau=z_1/z_2$.
Applying the conformally transformed boundary condition, 
we obtain an implicit equation for $\mu_n=z_2 m_n$ as the roots of 
\begin{equation}\label{eq:scalareig}
F(m_n)= J_{2/5} (\mu_n \tau) Y_{2/5} (\mu_n) - J_{2/5} (\mu_n) Y_{2/5} (\mu_n
\tau)=0.
\end{equation}
The $\zeta$-function is
\begin{equation}
\zeta(s)= n_V \int d^4 x \int \frac{d^4k}{(2\pi)^4} \sum_n \left(
\frac{k^2+m_n^2}{\mu_R^2}\right)^{-s}.
\end{equation}
The factor of $n_V$ is included since we have $n_V$ scalar fields.
Performing the $k$ integrals first gives us
\begin{equation}\label{eq:scalarzeta}
\zeta(s)= n_V \mu_R^{2s} \int d^4 x \frac{\Gamma(s-2)}{(4\pi)^2\Gamma(s)} 
\sum_n \mu_n^{4-2s} z_2^{2s-4}.
\end{equation}
The evaluation of the sum over $\mu_n$ is complicated since we only know
them through an implicit equation.
We leave the details of the evaluation of the sum in (\ref{eq:scalarzeta}) to
Appendix~\ref{sec:sum}.
Here, we simply quote the
result for the effective action of the conformally transformed operator;
\begin{equation}\label{eq:scalarea}
W^{\text{scalar}}_0 =  \frac{n_V}{(4\pi)^2} \int d^4 x
\left[\frac{A_{2/5}}{z_1^4} + 
\frac{B_{2/5}}{z_2^4} 
+ \frac{G_{2/5}(\tau)}{z_2^4} + \frac{2781}{80000}
\left( \frac{\ln(\mu_R z_1)}{z_1^4} + \frac{\ln(\mu_R z_2)}{z_2^4} \right)
\right].
\end{equation} 
The function $G_\nu(\tau)$ is
\begin{equation}
G_\nu(\tau)=\int_0^{\infty}dx x^3 
\ln \left(1-\frac{K_\nu(x)I_\nu(\tau x)}{I_\nu(x)K_\nu(\tau x)}\right),
\end{equation}
and the constants $A_\nu$ and $B_\nu$ are defined in Appendix~\ref{sec:sum}.
To obtain the full expression for the effective action, we must add the
cocycle function to this result. After a simple but lengthy calculation, the
details of which can be found in Appendix~\ref{sec:cocycle},
the cocycle function for this operator is found to be
\begin{equation}
C^{\text{scalar}}[\Omega]= \frac{n_V}{(4\pi)^2} \int d^4 x \left[
\frac{2781}{400000}\left(\frac{\ln (c z_1)}{z_1^4} 
+ \frac{\ln (c z_2)}{z_2^4} \right)+\frac{4601}{9600000}
\left(\frac{1}{z_1^4}+\frac{1}{z_2^4} \right) \right].
\end{equation}
The constant $c=6b_0/(5\sqrt2\alpha a_0)$ can be
absorbed into the renormalization scale $\mu_R$. After this redefinition 
we find the total effective action for the scalar field
\begin{eqnarray}
W^{\text{scalar}} = \frac{n_V}{(4\pi)^2} \int d^4 x &&
\left[\frac{A_{2/5}}{z_1^4} + \frac{B_{2/5}}{z_2^4} +
\frac{G_{2/5}(\tau)}{z_2^4} \right. \nonumber \\ 
&&\left. + \frac{8343}{200000}
\left( \frac{\ln(\mu_R z_1)}{z_1^4} + \frac{\ln(\mu_R z_2)}{z_2^4} \right)+
\frac{4601}{9600000}\left(\frac{1}{z_1^4}+\frac{1}{z_2^4}\right)
\right].
\end{eqnarray}

\subsection{The vector field}\label{sec:vector}

The five dimensional $M$-theory action contains $h^{1,1}$ $U(1)$ gauge fields
${\cal A}^i_\alpha$.
One of these gauge fields is the graviphoton ${\cal A}_\alpha$ of the gravity
multiplet. This
leaves $n_V$ gauge fields for the $n_V$ vector multiplets. After
decomposing the ${\cal A}^i_\alpha$ into the graviphoton and vector multiplet
gauge field by ${\cal A}^x_\alpha=b^x_i {\cal A}^i_\alpha$ and 
${\cal A}_\alpha=\frac23 b_i {\cal A}^i_\alpha$, the action for the vector
multiplet
gauge field to quadratic order is
\begin{equation}
S_{\textrm{gauge}}=\int_{\cal M}\! d \mu \ \frac 14 {\cal F}^x_{AB} 
{\cal F}_x^{AB}
\end{equation}
Integration by parts yields
\begin{equation}
S_{\textrm{gauge}}=\int_{\cal M} \! d\mu \ \frac 12 {\cal A}_x^A
\left( -\nabla^2 \delta_A^B + R_A^{\ B}+ \nabla_A \nabla^B \right) {\cal A}^x_B
\end{equation}
were we have neglected boundary terms. We work
in a tetrad frame. The covariant derivative contains the
connection 
$\nabla_\alpha=\partial_\alpha+i\omega_\alpha^{\ CD}\Sigma_{CD}$ where
$\left(\Sigma_C^{\ D}\right)_A^{\
B}=-\frac{i}{2}\left(\eta_{CA}\eta^{DB}-\delta_C^B\delta_A^D\right)$ is the 
generator of Lorenz transformations appropriate for a 
vector representation of the Lorenz group. Again, since the index
structure in the index $x$ is trivial, we drop this index from now on.

We can add to the action
a gauge fixing term $S_\textrm{gf}$ and anticommuting ghost fields $c,\bar{c}$
so that
$S_{\textrm{total}}=S_{\textrm{gauge}} + S_{\textrm{gf}} + S_{\textrm{ghost}}$
where
\begin{eqnarray}
S_{\textrm{gf}} &=&\int_{\cal M} \!d \mu \ \frac 12 F^2,\\
S_{\textrm{ghost}}& =&\int_{\cal M} \! d \mu\ {\bar c} 
\frac{\delta F}{\delta \Lambda} c .
\end{eqnarray}
A convenient choice of gauge fixing term will turn out to be
\begin{equation}\label{eq:gaugechoice}
F = \nabla \cdot {\cal A} +2 \omega_{;\alpha} {\cal A}^\alpha.
\end{equation}
Since $\phi_{;\alpha}=-6\omega_{; \alpha}$, this gauge fixing term can
also be expressed in terms of the background dilaton field.
The gauge fixed action for the gauge field
$S_\textrm{g+gf}=S_{\textrm{gauge}}+S_{\textrm{gf}}$
can then be written, after an integration by parts, as
\begin{equation}
S_{\textrm{g+gf}}=
\int_{\cal M} \! d \mu \ \frac 12 {\cal A}^A
\left( -\nabla^2 \delta_A^B + R_A^{\  B} +2
\omega_{;A}\nabla^B-2 \omega^{;B}\nabla_A-2\omega_{;A}^{\ B}+
4 \omega_{;A}\omega^{;B}\right) {\cal A}_B.
\end{equation}
The action can be simplified by introducing the derivative 
$(D_A)_B^{\ C}=(\nabla_A)_B^{\ C}+i (A_A)_B^{\ C}$, 
where the connection $A_A$ is chosen to be
\begin{equation}
(A_A)_B^{\ C} = -i\left( \omega^{;C}\eta_{AB} - \omega_{;B} \delta_A^C
\right) = 2 \omega_{;D}(\Sigma_A^{\ D})_B^{\ C}
\end{equation}
The gauge fixed operator can now be written in the form of a Laplace-type
operator, i.e.,
\begin{equation}\label{eq:gaugeop}
{\Delta}_A^{\  B} = - (D^2)_A^{\  B} + X_A^{\  B},
\end{equation}
where
\begin{equation}
X_A^{\  B} = R_A^{\  B} -2 \omega_{;A}^{\ \ B} + 
\omega_{;A}\omega^{;B}-\omega_{;C}\omega^{;C}\delta_A^B.
\end{equation}

Boundary conditions can be found from the requirement of invariance under BRST
transformations \cite{Moss:1997ip}. This ensures that the path integral is
gauge independent.
The two admissible sets of boundary conditions are known as
absolute and relative. Absolute boundary conditions have
\begin{equation}\label{eq:absbcs}
{\cal A}_N=0, 
\quad \left(D_N \delta_I^J + K_I^{\ J} \right){\cal
A}_J=0, \quad \nabla_N c=\nabla_N {\bar c} = 0\quad \textrm{on} \quad \partial
{\cal M}
\end{equation}
while relative boundary conditions are obtained by setting the gauge fixing
term
$F=0$ on the boundary, this leads to
\begin{equation}\label{eq:relbcs}
{\cal A}_J=0, 
\quad \left(D_N + K +2\omega_{;N} \right){\cal
A}_N=0, \quad c={\bar c} = 0\quad \textrm{on} \quad \partial {\cal M}.
\end{equation}
Although we have not written the boundary terms in the action explicitly, 
both these sets of boundary conditions cause the boundary terms generated
by the integration by parts 
to vanish. The choice of boundary condition is determined by the $Z_2$ symmetry
in the ``upstairs'' picture, which requires the tangential components $A_I$
to be odd under the $Z_2$ symmetry, and the $A_N$ to be even. In the
``downstairs'' picture, the boundary conditions can be derived from
the boundary conditions on the eleven dimensional three
form \cite{Moss:2003bk}. 
This fixes the boundary conditions to be relative -- i.e., the boundary
conditions are
given by Eq. (\ref{eq:relbcs}).

We now perform a conformal transformation and
calculate the eigenvalues of the conformally transformed operator. We
transform the operator and the boundary conditions according to 
(\ref{eq:xtrans}) and (\ref{eq:strans}) respectively. 
The reason for our choice of gauge fixing term (\ref{eq:gaugechoice})
should now become clear. After a conformal transformation, the 
connection $\tilde{A}_{A}=0$. 
\footnote{If we
had chosen the conventional gauge fixing term $F= \nabla \cdot {\cal A}$,
then the conformal transformation would introduce a connection into the
derivative which would couple the equations for the normal and tangential
components of the gauge field. The resulting system of equations does not seem
to be solvable in terms of elementary functions}
 
The one loop effective action can now be found from the
eigenvalues of the conformally transformed operator
$\Delta^\text{gfixed}_0$.
For the components of the gauge field tangential to the boundary, 
we find the solutions
\begin{equation}
\tilde{\cal A}_I = \sqrt{z}e^{i k_\mu x^\mu}(C_I J_{2/5}(m_n z) + D_I
Y_{2/5}(m_n z)),
\end{equation}
The tangential components obey Dirichlet boundary conditions, so the implicit
equation for the eigenvalues $\mu_n$ is 
\begin{equation}\label{eq:kkvec}
F(\mu_n)= J_{2/5}(\mu_n \tau)Y_{2/5}(\mu_n)-J_{2/5}(\mu_n)Y_{2/5}(\mu_n\tau)=0.
\end{equation}
The bulk solutions for the component of the gauge field in the orbifold
direction are
\begin{equation}
\tilde{\cal A}_{11} = \sqrt{z} e^{i k_\mu x^\mu}(C J_{3/5}(m_n z) + D
Y_{3/5}(m_n z)).
\end{equation}
The appropriate boundary conditions are Robin and therefore leads to an
implicit
equation for the eigenvalues involving both Bessel functions and their
derivatives. However, the particular combination in this case again reduces to
(\ref{eq:kkvec}).

The eigenvalues for the conformally transformed
gauge field operator are the same as for the scalar field case considered
in Section~\ref{sec:scalar}. Hence,
the effective action of the conformally transformed operator is the same as 
in (\ref{eq:scalarea}), except here we have an additional degeneracy factor of
5 for the five degrees of freedom of the gauge field.

The operator for ghost fields $c$ and $\bar{c}$,
\begin{equation}
\Delta^{\text{ghost}}=\frac{\delta  F}{\delta \Lambda} = \nabla^2 +2
\omega_{;\alpha} 
\nabla^\alpha
\end{equation}
is not yet in the canonical form of (\ref{eq:laplace}). However,
we can introduce a connection into the derivative to get the operator into
the appropriate form. We set $A_\alpha = -i \omega_{;\alpha}$, 
so that 
\begin{equation}
\Delta^{\text{ghost}}=\frac{\delta  F}{\delta \Lambda} = (\nabla+iA)^2
-\omega_{;\alpha}
\omega^{;\alpha} - \omega_{;\alpha}^{\ \ \alpha}.
\end{equation}
The eigenfunctions of the conformally rescaled operator are
\begin{equation}
\tilde{c}(z)= z^{7/10} e^{i k_\mu x^\mu}(A J_{2/5}(m_n z) + B Y_{2/5}(m_n z)).
\end{equation}
The ghost fields obey Dirichlet boundary conditions, hence the implicit
equation for $\mu_n$ is again
\begin{equation}
F(\mu_n)= J_{2/5}(\mu_n\tau)Y_{2/5}(\mu_n)-J_{2/5}(\mu_n)Y_{2/5}(\mu_n \tau)=0.
\end{equation}
The $\zeta$-function for the ghost operator
follows from the scalar case in Section~\ref{sec:scalar}, save for
an additional degeneracy factor of 2.
We can see immediately that the ghosts will cancel the contribution from
two components of the gauge field, leaving us with 3 physical degrees of
freedom
as we would expect.

The calculation of the cocycle function turns out to be relatively simple in
this 
gauge. More detail can be found in Appendix~\ref{sec:cocycle}.
We find the cocycle function to be (including the ghost fields)
\begin{equation}
C^{\text{gauge}}=\frac{n_V}{(4 \pi)^2}\int d^4 x 
\left[\frac{8343}{40000}\left(\frac{\ln (c z_1)}{z_1^4}
+\frac{\ln(c z_2)}{z_2^4}
\right)
-\frac{193669}{9600000}\left(\frac{1}{z_1^4}+\frac{1}{z_2^4}\right)\right].
\end{equation}
The total effective action for the gauge fields
$W^{\text{gauge}}=W^{\text{gauge}}_0+C^{\text{gauge}}$ is
\begin{eqnarray}
W^{\text{gauge}}= \frac{3 n_V}{(4\pi)^2} \int d^4 x
&&\left[\frac{A_{2/5}}{z_1^4} + 
\frac{B_{2/5}}{z_2^4} 
+ \frac{G_{2/5}(\tau)}{z_2^4} \right. \nonumber \\
&&\left.+ \frac{8343}{200000}
\left( \frac{\ln(\mu_R z_1)}{z_1^4} + \frac{\ln(\mu_R z_2)}{z_2^4} \right)
-\frac{193669}{28800000}\left(\frac{1}{z_1^4}+\frac{1}{z_2^4}\right)\right].
\end{eqnarray}

We have checked the gauge independence of this result by calculating
various quantities in different gauges. Firstly, we can evaluate
$\zeta(0)$ via the $a_{5/2}(1,\Delta)$ heat kernel coefficient using the
Lorentz gauge $F=\nabla \cdot {\cal A}$. We have also performed the calculation
of the flat space effective
action in an axial-type gauge with gauge fixing term 
$F=\omega_{;A} {\cal A}^A$. In all three gauges we find exact
agreement between the various quantities.

\subsection{The fermion field}\label{sec:fermion}

The vector multiplet contains $n_V$ pairs of symplectic real fermions 
$\lambda^x_a$. The symplectic index $a=1,2$ is raised and lowered using an
antisymmetric metric $\epsilon_{ab}$.
The fermion action in the vector multiplet is
\begin{equation}
S_\textrm{fermion}=\int_{\cal M} \! d \mu \ 
\left( \frac12 \bar{\lambda}^{a\, x} \gamma^\alpha
\nabla_\alpha \lambda_{a\, x}
- \frac12 M_{a\ x}^{\ b \ y} \bar{\lambda}^{a\, x} \lambda_{b\, y} \right)
\end{equation}
The mass matrix $M_{a\ x}^{\ b \ y}$ depends on the background dilaton field.
We use the fermion mass term for gauged supergravity models
\cite{Gunaydin:1985}.
(We believe that the mass term quoted in \cite{Lukas:1998tt} is incorrect). 
Explicitly, in this background,
\begin{equation}
M^a_{\ \, b \, xy} = 
-\alpha \frac{\sqrt{2}}{12}e^{-\Phi}{(\sigma_3)}^a_{\ b} g_{xy}
\end{equation}
where ${(\sigma_3)}^a_{\ b}=\textrm{diag}(1,-1)$ is the Pauli matrix.
Again, the operator is diagonal in indices $x$, so we leave them out from
now on for simplicity. 
Boundary conditions on the fermions are
\begin{equation}\label{eq:fermionbound}
\gamma^{11} \lambda^a = {(\sigma_3)}^a_{\ b} \lambda^b \quad \textrm{on}
\quad \partial {\cal M}
\end{equation}

We define a column spinor
\begin{equation}
\Lambda=\left(\begin{array}{c}\lambda^1 \\ \lambda^2 \end{array} \right)
\end{equation}
and matrices
\begin{equation}
\Gamma_A=\left(\begin{array}{cc}
                \gamma_A & 0 \\
		    0     & -\gamma_A
		\end{array} \right).
\end{equation}
We can now write the action as
\begin{equation}
S_\textrm{fermion}=\int_{\cal M} \! d \mu \ \bar{\Lambda} \left(\Gamma^A
\nabla_A
- m \right) \Lambda,
\end{equation}
where $m=-\alpha \frac{\sqrt{2}}{12}e^{-\Phi}$ and we have defined
$\bar{\Lambda}={\Lambda}^\dagger \Gamma^0$. 
Introducing the projection operators 
\begin{equation}
P_\pm = \frac12 (1\pm \Gamma^{11}),
\end{equation}
the boundary conditions on the fermions (\ref{eq:fermionbound}) can be written
\begin{equation}
P_- \chi =0 \quad \textrm{on}\quad \partial {\cal M}.
\end{equation}

The Dirac equation ${\cal D}=\slash \nabla - m$ is a first order
differential 
equation and hence not
of Laplace type (\ref{eq:laplace}). However, we can form a
second order differential operator from the
square of the Dirac operator. We make two copies of the spin bundle with
opposite orientations, $S^+$ and $S^-$. The Dirac operator maps spinors in
$S^+$, denoted $\Lambda$, to image spinors in $S^-$, denoted $\Lambda'$. 
The adjoint operator is
${\cal D}^*=-\slash \nabla -m$. The eigenvalue equation
is then
\begin{equation}
{\cal D} \Lambda = \eta \Lambda', \quad {\cal D}^* \Lambda' = \eta
\Lambda,
\end{equation}
with the boundary conditions
\begin{equation}
P_- \Lambda =0, \quad P_+ \Lambda'=0 \quad \text{on} \quad \partial {\cal M}.
\end{equation}

Provided $\eta \neq 0$, we can form a second order operator
with eigenvalue $\eta^2$. One choice of second order operator is
\begin{equation}\label{eq:delta}
\Delta={\cal D}^* {\cal D}=-\nabla^2+\frac14 R + m^2 + \slash \partial m
\end{equation}
where we have allowed for the possibility that $m$ is non-constant
across the orbifold. The operator $\Delta$ maps spinors in $S^+$ to $S^+$.
The boundary conditions on the fermions are
\begin{equation}\label{eq:flb}
P_- \Lambda =0, \quad P_+ {\cal D} \Lambda=0, \quad \textrm{on}
\quad \partial {\cal M}.
\end{equation}
There is also an alternative squared operator
\begin{equation}\label{eq:deltahat}
\Delta^*={\cal D} {\cal D}^*=-\nabla^2+\frac14 R + m^2 - \slash \partial m
\end{equation}
which maps spinors in $S^-$ to $S^-$. The boundary conditions are
\begin{equation}\label{eq:flb2}
P_+ \Lambda =0, \quad P_- {\cal D}^* \Lambda=0, \quad \textrm{on}
\quad \partial {\cal M}.
\end{equation}
The spectrum of (\ref{eq:delta}) and (\ref{eq:deltahat}) are identical if there
are no zero modes.

The massless Dirac operator transforms covariantly under the conformal 
transformation (\ref{eq:contrans}),
\begin{equation}
\tilde{\slash \nabla} = 
\Omega^{-3} \slash \nabla \Omega^{2}.
\end{equation}
The presence of a mass breaks the conformal invariance. We can, however,
introduce a mass that scales under the conformal transformation, such that
$\tilde{m}=\Omega^{-1} m$. Then both $\cal D$ and ${\cal D}^*$ transform
covariantly under
conformal transformations.
The boundary conditions (\ref{eq:flb}) and (\ref{eq:flb2})
are invariant under conformal transformations. 
The squared Dirac operator does not transform covariantly. However, in appendix
B it is shown that the effective action for fermions can be written in terms
of the effective action in the conformally related space using
\begin{equation}
W=W_0-\int_0^1 d\sigma \ \frac12
\left(a_{5/2}(\omega,\Delta)+a_{5/2}(\omega,\Delta^*)
\right)\label{fcocycle}
\end{equation}
As we have mentioned, if there are no zero modes, then the spectrum of $\Delta$
and $\Delta^*$
are equivalent and $a_{5/2}(\omega,\Delta)=a_{5/2}(\omega,\Delta^*)$.

We are therefore lead to consider
the eigenvalues of the operator $\Delta_0 = {\cal D}_0^* {\cal D}_0$ in flat
space. In this case, the bulk solutions are again related to the
Bessel functions. Defining integration constants $A^{(\pm)}$ and $B^{(\pm)}$
such that $\Gamma^{11} A^{(\pm)}=\pm A^{(\pm)}$ and 
$\Gamma^{11} B^{(\pm)}=\pm B^{(\pm)}$, we can write the solution as
\begin{equation}
\tilde{\Lambda}_n(z)=\sqrt{z}\left(A^{(-)} J_{2/5} (m_n z) +
B^{(-)} Y_{2/5} (m_n z) + 
A^{(+)} J_{3/5} (m_n z)+ 
B^{(+)} Y_{3/5} (m_n z)\right).
\end{equation} 
The boundary conditions (\ref{eq:flb}) in the conformally 
transformed space can be written as
\begin{equation}\label{eq:fermiconformbcs}
P_- \tilde{\Lambda} =0, \quad \left(\partial_z -
m_0\right)P_+\tilde{\Lambda}=0, \quad \textrm{on}
\quad \partial {\cal M}.
\end{equation}
This gives us the implicit equation for the eigenvalues
$\mu_n$; 
\begin{equation}\label{eq:feig}
F(\mu_n)= J_{2/5} (\mu_n \tau) Y_{2/5} (\mu_n) - 
J_{2/5} (\mu_n) Y_{2/5} (\mu_n \tau)=0.
\end{equation}

The fermion $\zeta$-function is
\begin{equation}
\zeta(s)= 4 n_V \mu_R^{-2s} \int d^4 x \frac{\Gamma(s-2)}{(4\pi)^2\Gamma(s)} 
\sum_n \mu_n^{4-2s} z_2^{2s-4}.
\end{equation}
Here, the degeneracy factor of $4$ is equal to the number of degrees of freedom
of
the fermion field. The sum over eigenvalues can be evaluated
using results from Appendix~\ref{sec:sum} to give
\begin{equation}
W^\text{fermion}_0 = -n_V \frac{4}{(4\pi)^2} \int d^4 x \left[
\frac{A_{2/5}}{z_1^4} + 
\frac{B_{2/5}}{z_2^4} 
+ \frac{G_{2/5}(\tau)}
{z_2^4} + \frac{2781}{80000}
\left( \frac{\ln(\mu_R z_1)}{z_1^4} + \frac{\ln(\mu_R z_2)}{z_2^4} \right)\right].
\end{equation}

The cocycle function for this operator can be calculated using
results from Appendix~\ref{sec:cocycle} to be
\begin{equation}
C^\text{fermion}[\Omega]=-4\frac{n_V}{(4\pi)^2}\int d^4 x 
\left[\frac{2781}{400000}\left(\frac{\ln(cz_1)}{z_1^4}
+\frac{\ln(c z_2)}{z_2^4}
\right)
-\frac{18129}{3200000}\left(\frac{1}{z_1^4}+\frac{1}{z_2^4}\right)\right].
\end{equation}
The effective action for the fermion in the vector multiplet is then
\begin{eqnarray}
W^\text{fermion} = -n_V \frac{4}{(4\pi)^2} \int d^4 x &&\left[
\frac{A_{2/5}}{z_1^4} + 
\frac{B_{2/5}}{z_2^4} 
+ \frac{G_{2/5}(\tau)}{z_2^4} \right. \nonumber \\
&& \left. + \frac{8343}{200000}
\left( \frac{\ln(\mu_R z_1)}{z_1^4} + \frac{\ln(\mu_R z_2)}{z_2^4} \right) -
\frac{18129}{3200000}\left(\frac{1}{z_1^4}+\frac{1}{z_2^4}\right)\right].
\end{eqnarray}

\subsection{Total effective action}

The effective actions for all the fields in the vector multiplets contain the
function $G_\nu(\tau)$. We can approximate this function
for small brane separation $\tau \rightarrow 1$ using the
asymptotic expansion of the Bessel function as \cite{Garriga:2001ar}
\begin{equation}
z_2^{-4} G_\nu(\tau) \approx -\frac{3}{8} \frac{\zeta_R(5)}{{(z_1-z_2)}^4}, \quad 1-\tau
\ll 1
\end{equation}
Hence, the effective actions of the individual fields diverge
as the brane separation goes to zero. 

We can see from the results in Section~\ref{sec:scalar},
\ref{sec:vector} and \ref{sec:fermion} that the contributions to the
effective action from the flat space effective potential $W_0$ will
cancel between the bosonic and fermionic sectors of the multiplet.
The logarithmic terms in the cocycle functions also cancel. There
is, however, a piece left over from the cocycle function, taking the
form
\begin{equation}
W^{\textrm{total}} = W^{\text{scalar}}+W^{\text{gauge}}+W^{\text{fermion}}
=  \frac{n_V}{(4\pi)^2} \int d^4\! x \ \frac{89}{30000}\left(\frac{1}{z_1^4}+\frac{1}{z_2^4}
\right)
\end{equation}
Although the individual bosonic and fermionic contributions to the effective
potential
diverge when the separation of the two branes is reduced to zero, the finite
residual term in the total effective action remains finite. From (\ref{eq:z1}), it can be seen that the residual term in the effective action
is of order $\alpha^4$ in the brane charge expansion.

In the low energy effective four dimensional theory the moduli $z_1$ and $z_2$
contribute to the dilaton $S$ and axion $T$ moduli fields and we would not
expect to find an explicit superpotential unless the supersymmetry was broken.
One way in which
the apparent contradiction with our result may be resolved would be through
higher
derivative terms in the Ho\v rava-Witten model. Curvature terms related to
string corrections have been discussed in \cite{Lukas:1998ew}. As an example,
consider
\begin{equation}
S_{R^5} = \frac{1}{16\pi \kappa^2} {\left(\frac{\kappa^{2/3}}{4 \pi}\right)}^3
\int_{\partial {\cal M}} \sqrt{g} d^{10} \!x \  I_3(R) I_2(R),
\end{equation} 
where $I_3(R)$ is the cubic combination of curvature terms which appear in the
six dimensional Gauss-Bonnet identity and $I_2(R)$ is any quadratic combination
of curvature tensors. When reduced to five dimensions, the action would contain
a term
\begin{equation}
S_{R^2}=\chi \int_{\partial {\cal M}} \sqrt{g} d^4 \! x I_2(R)
\end{equation}
where $\chi$ is the Euler number of the Calabi-Yau manifold used in the
reduction
to five dimensions. Since $\chi=2(h^{1,1}-h^{1,2})$, this term has the correct
dependence on $n_V$, $z_1$ and  $z_2$
to be able to cancel the one loop term from the vector multiplets. 

\section{Conclusions}

We have calculated the one-loop effective action for the vector multiplets
related to the $h^{1,1}$ moduli of the Calabi-Yau 
compactification of heterotic $M$-theory. We find that effective action of the
individual fields induces a potential for the moduli which diverges as the
brane separation is reduced. These terms cancel between the fermions and
bosons in the vector multiplet. However, boundary terms do not cancel
exactly. We propose that these boundary terms could be cancelled by
higher derivative terms in the $M$-theory action. If this is the case, then
this term should be taken into account in future work on the one-loop
effective action in heterotic $M$-theory. For, example, it would be interesting
to consider the effect of supersymmetry breaking by gaugino condensation
\cite{Horava:1996vs, Lukas:1998rb} on the quantum effective action. Since
supersymmetry is broken by the topology of the orbifold direction, one would
expect that the casimir effect would be non-zero and give an extra contribution
to the classical moduli potential induced by the gaugino condensate. 

\acknowledgments

We are grateful to James Gray for discussions about heterotic $M$-theory.

\appendix

\section{Evaluation of the $\zeta$-function}\label{sec:sum}

\begin{figure}[ht]
\psfragscanon
\psfrag{cc}{${\cal C}$}
\includegraphics[height=4in]{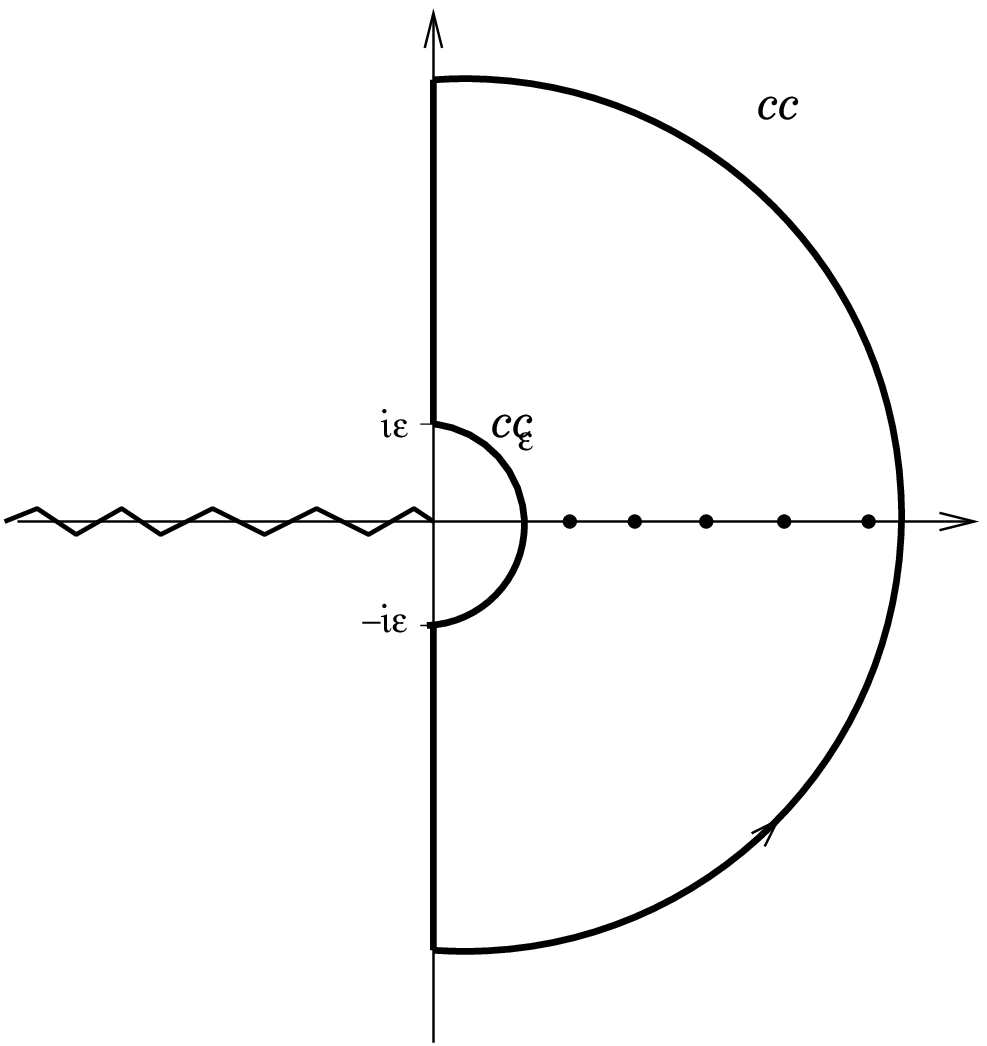}
\label{fig:contour}\caption{Contour used for the contour integral in
(\ref{eq:zetaint})}
\end{figure}

The $\zeta$-function is
\begin{equation}
\zeta(s)= \int d^4 x \int \frac{d^4k}{(2\pi)^4} \sum_n \left(
\frac{k^2+m_n^2}{\mu_R^2}\right)^{-s}.
\end{equation}
Performing the momentum integrals gives us
\begin{equation}\label{eq:zeta}
\zeta(s)= \mu_R^{2s} \int d^4 x \frac{\Gamma(s-2)}{(4\pi)^2\Gamma(s)} 
\sum_n \mu_n^{4-2s} z_2^{2s-4}
\end{equation}
The summation of the masses is complicated since we only know the masses as
implicit solutions of an equation $F(\mu_n)=0$. However, we can follow 
\cite{Elizalde:1993,Leseduarte:1996av,Bordag:1996gm}
and use the residue theorem to write the sum as a contour integral, that is
\begin{equation}\label{eq:zetaint}
\hat{\zeta}(2s-4)=\sum_n \mu_n^{4-2s} 
= \int_{\cal C} dz \ z^{4-2s} \frac{d}{dz} \ln |F(z)|.
\end{equation}
Here the contour ${\cal C}$ is any contour which encloses the positive zeros
of $F(z)$. For the implicit eigenvalue equation (\ref{eq:scalareig}), we must
restrict $s$
to lie in the range $5/2<s<3$.
We chose to work with the contour shown in Figure~\ref{fig:contour}.
For $s>\frac52$, the contribution to the integral from the large
semi circle vanish, and we are left with the contribution along the imaginary
axis and the small semi circle.
This results in
\begin{equation}
\hat{\zeta}(2s-4) = \frac{\sin(\pi s)}{\pi} 
\int_\epsilon^\infty dx \ x^{4-2s} \frac{d}{dx} \ln \left|P(x)\right| 
+ \int_{{\cal C}_\epsilon}\frac{dz}{2\pi i} z^{4-2s} \frac{d}{dz} \ln |F(z)|,
\end{equation}
where we have defined $P(x)=F(ix)$. 
Explicitly, we take (up to an irrelevant multiplicative constant),
\begin{equation}
P(x)=I_\nu(\tau x)K_\nu(x)- I_\nu(x) K_\nu(\tau x).
\end{equation}
We also define
\begin{equation}
P_a(x) = I_\nu(x)K_\nu(\tau x).
\end{equation}
We can now write
\begin{equation}\label{eq:divide}
\int_\epsilon^\infty dx \ x^{4-2s} \frac{d}{dx} \ln \left|P(x)\right|
 =\int_\epsilon^\infty dx \ x^{4-2s} 
  \frac{d}{dx} \ln \left|\frac{P(x)}{P_a(x)}\right|
+ \int_\epsilon^\infty dx \ x^{4-2s} 
  \frac{d}{dx} \ln \left|P_a(x)\right|.
\end{equation}
The first term on the RHS of (\ref{eq:divide}) is now analytic at $s=0$. 
The only obstruction to analytically continuing to $s=0$ is the second term.
We can write
\begin{equation}
I_\nu(x)=\frac{e^x}{\sqrt{2\pi x}} \Sigma^I_\nu(x), \quad  
K_\nu(x)=\sqrt{\frac{\pi}{2x}} e^{-x} \Sigma^K_\nu(x).
\end{equation}
Using the asymptotic expansion of Bessel functions, we can define
constants $\beta_n$ by
\begin{equation}\label{eq:lnsigmai} 
\ln \left|\Sigma_\nu^{I}\right| = \sum_{n=1}^\infty \beta_n x^{-n}.
\end{equation}
Explicit expressions for the $\beta_n$ can be found in \cite{Flachi:2001pq}.
We also have that
\begin{equation}\label{eq:lnsigmak}
\Sigma^{I}_\nu (x) \simeq \Sigma^{K}_\nu (-x).
\end{equation}
We define regularised functions by subtracting off the terms which cause
the integrand to diverge at large $x$,
\begin{subequations}
\label{eq:IK}
\begin{eqnarray}
{\cal U}_I (x)&=&
\frac{d}{dx} \ln \left|\Sigma_\nu^I (x)\right|
+\sum_{n=0}^3 n \beta_n x^{-n-1} + 4 \beta_4 x^{-5} e^{-k/x}\\
{\cal U}_K (x)&=&
\frac{d}{dx} \ln \left|\Sigma_\nu^K (x)\right|
+\sum_{n=0}^3 (-1)^n n \beta_n x^{-n-1} + 4 \beta_4 x^{-5} e^{-k/x}.
\end{eqnarray}
\end{subequations}
We can now write the second term on the RHS of (\ref{eq:divide}) 
using \ref{eq:IK} ,\ref{eq:lnsigmai} and \ref{eq:lnsigmak}. The integrals of
the
terms which we have subtracted off can be performed for $s>1$.
We can now analytically continue to $s<\frac12$ and take the limit 
$\epsilon=0$. This gives
\begin{equation}
\hat{\zeta}(2s-4) = -\frac{4\sin \pi s}{\pi} \left\{g_\nu(s) 
+b_\nu(s) + a_\nu(s) \tau^{2s-4} +\beta_4 k^{-2s} \Gamma (2s)(1+\tau^{2s-4})
\right\}, 
\end{equation}
where we have defined the functions
\begin{equation}
g_\nu(s)=-\frac14\int_0^\infty dx \ x^{4-2s}
\frac{d}{dx} \ln \left|\frac{P(x)}{P_a(x)}\right|,
\end{equation}
\begin{equation}
b_\nu(s)=-\frac14\int_0^\infty dx\ x^{4-2s} {\cal U}_I (x)
\end{equation}
and
\begin{equation}
a_\nu(s)=-\frac14\int_0^\infty dx\ x^{4-2s} {\cal U}_K (x)
\end{equation}
At $s=0$, we have, after a redefinition of the renormalization constant $\mu_R$,
\begin{equation}
\zeta(0) = -\frac{\beta_4}{16 \pi^2} \left(z_1^{-4}+z_1^{-4}\right)
\end{equation}
\begin{equation}
\zeta'(0) = -\frac{1}{8 \pi^2} \left( \frac{G_\nu}{z_2^4}
+ \frac{B_\nu}{z_2^4} + \frac{A_\nu}{z_1^4}\right)
-\frac{\beta_4}{8\pi^2}\left(
\frac{\ln z_1\mu_R}{z_1^4} + \frac{\ln z_2\mu_R}{z_2^4} \right),
\end{equation}
where $B_\nu=b_\nu(0)$, $A_\nu=a_\nu(0)$, and $G_\nu=g_\nu(0)$.

\section{Cocycle functions for spin 0,1/2 and 1}\label{sec:cocycle}

We begin by reviewing the relationship between the cocycle functions and the
heat kernel coefficients. Suppose that $\Delta(\sigma)$ is a one parameter
family of operators
\begin{equation}
\Delta(\sigma)=e^{-2\sigma\omega}\Delta_0
\end{equation}
Consider the derivative of the generalised $\zeta$ function
\begin{equation}
\partial_\sigma\zeta(s)=-2s\,{\rm tr}(\Delta^{-s}\omega).
\end{equation}
We denote this trace by $\zeta(s,\omega]$. The derivative of the effective
action
\begin{equation}
\partial_\sigma W=\zeta(0,\omega]=a_{5/2}(\omega,\Delta)
\end{equation}
where $a_{5/2}(\omega,\Delta)$ is a heat kernel coefficient. Integrating this
with respect to $\sigma$ gives equation (\ref{cocycle}).

In the case of fermions, we have an operator $\Delta={\cal D}^*{\cal D}$ where
\begin{equation}
{\cal D}\equiv{\cal D}(\sigma)=e^{-3\sigma\omega}{\cal D}_0e^{2\sigma\omega}
\end{equation}
This time
\begin{equation}
\partial_\sigma\Delta(\sigma)=
-\omega{\cal D}^*{\cal D}-{\cal D}^*\omega{\cal D}
-2[\omega,\Delta]
\end{equation}
Using the cyclic property of the trace, this results in
\begin{equation}
\partial_\sigma\zeta(s)=-s\,{\rm tr}(\Delta^{-s}\omega)
-s\,{\rm tr}({\cal D}\Delta^{-1-s}{\cal D}^*\omega)
\end{equation}
Note that ${\cal D}\Delta^n{\cal D}^*=\Delta^*{}^{n+1}$, and 
using the definition of $\Delta^{-s}$ from \cite[page 78]{Gilkey:1995mj}
we now find
\begin{equation}
\partial_\sigma W=-\frac12
\left(a_{5/2}(\omega,\Delta)+a_{5/2}(\omega,\Delta^*)\right).
\end{equation}
Following the steps above now gives equation (\ref{fcocycle}).

The $a_{5/2}$ heat kernel coefficient is known for general Laplace type
operators of the form
\begin{equation}
\Delta = -D^2 + X
\end{equation}
where $D=\nabla+iA$ \cite{Branson:1999jz}.
Its explicit form for a particular operator on a particular background is quite
lengthy. In this appendix we show some more detail in the calculation
of the cocycle function for spin 0,1/2 and 1 fields
in five dimensions for the case of a space which is conformal to flat space
with a flat boundary, which will result in expression which are much more
manageable that the full general
coefficient. 

The field is split up into components obeying Dirichlet boundary
conditions and components obeying Robin type boundary conditions by projection
operators, such that
\begin{equation}
P_- \phi = 0, \quad (D_N-{\cal S})P_+ \phi = 0 \quad \text{on} \quad \partial
{\cal M},
\end{equation}
where the projection operators obey $P_++P_-=1$, $P_\pm P_\pm = P_\pm$ and
$P_\pm P_\mp =0$. The combination $\chi=P_+-P_-$. The field strength of the
derivative is
defined as $\Omega_{AB}=[D_A,D_B]$. Tangential derivatives with respect to
the Levi-Civita connection of the boundary are denoted by ``$|$'' and 
contain the spin connection and the connection $A$. Since
the boundary is intrinsicly flat, this means 
$E_{|I}=\partial_I E+ i\omega_I^{\ CD}\Sigma_{CD} E+i A_I E$. 
For example, if $D$ is the covariant derivative for spin $1/2$ fields, then
 $\Gamma^J_{|I} = -K_I^{\ J} \Gamma^N$.  
 
The $a_{5/2}(f,\Delta)$ heat kernel coefficient 
can be written as
\begin{equation}
a_{5/2}(f,\Delta)= \frac{1}{5760} {(4\pi)}^{-2}
			\text{tr} \int_{\partial {\cal M}} d \mu \left\{f a^{(1)}
			+f_{;N} a^{(2)} + f_{;NN} a^{(3)}
			+f_{;NNN} a^{(4)} + f_{;NNNN} a^{(5)}\right\}
\end{equation}
where
in five dimensions
with a four dimensional {\it maximally symmetric} boundary, the $a^{(n)}$
simplify to
\begin{equation}
\begin{array}{lll}
a^{(1)} &= & 
			-360 \chi X_{;NN} 
			+1440 {\cal S} X_{;N} 
			+720 \chi X^2 
			-240\chi X R \\
&& 			+48 \chi R_{;AA}
			+20\chi R^2 
			-8 \chi R_{AB}R_{AB}
			+8\chi R_{ABCD}R_{ABCD} \\
&&			+120 \chi X R_{NN} 
			-20 \chi R R_{NN}
			+480 {\cal S}^2 R
			+12\chi R_{;NN}
			+15\chi R_{NN;NN} \\
&&			-270 {\cal S} R_{;N} 
			+120 R_{NN} {\cal S}^2
			+960 {\cal S} {\cal S}_{|II}
			+16 \chi R_{INJN}R_{IJ}
			-17\chi R_{NN} R_{NN}\\
&&			-10\chi R_{INJN} R_{INJN} 
			-2880 X {\cal S}^2
			+1440 {\cal S}^4
			+\left(90P_++450P_-\right) X_{;N} K \\
&&			-\left(\frac{111}{2}P_++42P_-\right)K R_{;N}
			-\frac{15}{2} P_+ K R_{NN;N}
			-1440K{\cal S} X 
			+90 {\cal S} K R_{NN} \\
&&			+225{\cal S} K R   
			+\left(-\frac{405}{2}P_++\frac{135}{2}P_-\right) X K^2 \\ 
&&			+\left(\frac{73}{2}P_+-\frac{71}{4} P_-\right) K^2 R 
			+\left(-\frac{313}{16}P_++\frac{433}{16}P_-\right)K^2 R_{NN} \\
&&			+270 {\cal S} K^3
			+1170 {\cal S}^2 K^2
			+2160 {\cal S}^3 K
			+\left(\frac{5619}{256}P_++\frac{51}{256}P_-\right) K^4 \\
&&			-180 X^2
			+180 \chi X \chi X 
			-120 {\cal S}_{|I} {\cal S}_{|I} 
			+720 \chi {\cal S}_{|I} {\cal S}_{|I} \\
&&			-\frac{105}{4} \Omega_{IJ} \Omega_{IJ} 
			+120 \chi \Omega_{IJ} \Omega_{IJ} 
			+\frac{105}{4} \chi \Omega_{IJ} \chi \Omega_{IJ}
			-45 \Omega_{IN}\Omega_{IN}  \\
&&			+180 \chi \Omega_{IN}\Omega_{IN}
			-45 \chi \Omega_{IN} \chi \Omega_{IN}
			+360 \left(\Omega_{IN} {\cal S}_{|I} \chi -\Omega_{IN} \chi {\cal
S}_{|I}\right) \\ 
&&			-90 \chi \chi_{|I} \Omega_{IN}
			-180 \chi_{|I} \chi_{|J} \Omega_{IJ} 
			+ 90 \chi \chi_{|I} \chi_{|J} \Omega_{IJ}
			+90\chi \chi_{|I} \Omega_{IN;N} \\
&&			+120 \chi \chi_{|I} \Omega_{IJ|J}
			-300 \chi_{|I} X_{|I}
			+180 \chi_{|I} \chi_{|I} X 
			+90 \chi \chi_{|I} \chi_{|I} X \\
&&			-240 \chi_{|II} X 
			-30 \chi_{|I} \chi_{|I} R
			-60 \chi_{|I} \chi_{|J} R_{IJ} 
			+30 \chi_{|I} \chi_{|J} R_{INJN} \\
&&			-\frac{1485}{32} \chi_{|I} \chi_{|J} K^2 
			-405 \chi_{|I} {\cal S}_{|I} K
			+\frac{15}{4} \chi_{|I} \chi_{|I}\chi_{|J} \chi_{|J} 
			+\frac{15}{8} \chi_{|I} \chi_{|J}\chi_{|I} \chi_{|J} \\
&&			-\frac{15}{4} \chi_{|II} \chi_{|JJ} 
			-\frac{105}{2} \chi_{|IJ} \chi_{|IJ} 
			-15 \chi_{|I} \chi_{|I} \chi_{|JJ}
			-\frac{135}{2}\chi_{|J} \chi_{|IIJ} \\
a^{(2)} &=&
 			 \left(\frac{195}{2}P_+-60P_-\right) R_{;N}
			-240 R{\cal S}
			+90 R_{NN} {\cal S}
			-270 {\cal S}_{|II} \\ 
&&			-\left(630P_+-450P_-\right) X_{;N}
			+1440 X {\cal S}
			-720 {\cal S}^3 \\
&&			+\left(90P_++450P_-\right) K X
			+\left(\frac{555}{16}P_+-\frac{75}{16}P_-\right) K R_{NN}
			-\left(\frac{75}{4}P_+ +\frac{135}{2} P_- \right) K R \\
&&			-600 K {\cal S}^2
			-\frac{1335}{8} K^2 {\cal S}
			-\left(\frac{969}{64}P_+-\frac{45}{64}P_-\right) K^3 \\
&&			+210 \chi_{|I} {\cal S}_{|I}
			+\frac{735}{32} \chi_{|I} \chi_{|I} K
			+135 \chi \chi_{|I} \Omega_{IN} \\
a^{(3)} &=& 		 60 \chi R 
			-90 \chi R_{NN} 
			-360 \chi X 
			+360 {\cal S}^2 \\
&&			+30 K {\cal S}
			-\left(\frac{15}{32}P_++\frac{1485}{32}P_-\right) K^2
			-30 \chi_{|I} \chi_{|I}  \\
a^{(4)} &=& 	 	-180 {\cal S} +\left(30P_+-105P_-\right) K \\
a^{(5)} &=& 		45\chi 
\end{array}
\end{equation}

The cocycle function for a conformally flat five dimensional metric 
\begin{equation}
ds^2=e^{-2 \omega(z)} \left( dx^{\mu} dx^{\nu} \eta^{(4)}_{\mu\nu}+dz^2 \right) 
\end{equation}
has the general form
\begin{equation}
C[\Omega] = \frac{1}{5760} (4\pi)^{-2} \int_{\partial{\cal M}}d \mu
\left(\omega A + B \right),
\end{equation}
where
\begin{eqnarray}
A &=& c_1 \omega '''' + c_2 \omega ''' \omega ' + c_3 \omega''^2 + c_4 \omega
'' \omega'^2 + c_5 \omega'^4, \\
B &=& d_1 \omega '''' + d_2 \omega ''' \omega ' + d_3 \omega''^2 + d_4 \omega
'' \omega'^2 + d_5 \omega'^4.
\end{eqnarray}
We have used a prime ($'$) to represent derivatives with respect to the
conformal coordinate $z$. In the following section we will
calculate the coefficients $c_n$ and $d_n$ for scalar, vector and fermion
fields in five dimensional conformally flat metrics
with maximally symmetric boundaries.

\subsection{Scalar fields}
\newcommand{\hz}{\hat{\xi}}
\newcommand{\he}{\hat{\eta}}

We restrict attention to operators with $X=\xi R$. The scalar field in the
vector multiplet considered in Section~\ref{sec:scalar}
has this form with $\xi=2/7$.

\subsubsection{Dirichlet boundary conditions}
For Dirichlet boundary conditions, the field $\phi$ vanishes on the boundary,
\begin{equation}
\phi=0 \quad \textrm{on} \quad \partial {\cal M}.
\end{equation} 
The projection operators are therefore simply
\begin{equation}
P_- =1, \quad P_+=0, \quad \chi = P_+-P_- = -1.
\end{equation}
Tangential derivatives of the projection operators are zero as is
the field strength of the connection,
\begin{equation}
\chi_{|I} =0, \quad \chi_{|IJ} =0, \quad \chi_{|IJK} =0,\quad \Omega_{AB}=0.
\end{equation}
The cocycle function for scalar fields with Dirichlet boundary conditions in
this geometry was calculated in \cite{Garriga:2001ar}
for the special case of when $\omega=-\beta \ln z$. We
find complete agreement with their results. The coefficients $c_{i}$ and
$d_{i}$ are listed in Table~\ref{tab:coeff}.

\subsubsection{Robin boundary conditions}

Robin boundary conditions have some combination of the field and its derivative
equal to zero on the boundary,
\begin{equation}
\left(D_N-{\cal S}\right)\phi=0 \quad \textrm{on} \quad \partial {\cal M}.
\end{equation} 
The projection operators are therefore
\begin{equation}
P_- =0 \quad P_+=1 \quad \chi = P_+-P_- = 1.
\end{equation}
Again, tangential derivatives of the projection operators are zero,
as is the commutator of two derivatives

We restrict our attention to boundary conditions with ${\cal S}=-\eta K$, where
$K$ is the
trace of the extrinsic curvature of the boundary. The scalar field in the
vector multiplet
we consider in Section~\ref{sec:scalar} is of this form with $\eta=1/2$. We
present the coefficients
$c_i$ and $d_i$ in Table~\ref{tab:coeff}.

\subsection{Vector Fields}

The gauge fixed operator in the gauge $F=\nabla \cdot {\cal A} +2\omega_{;A}
{\cal A}^{;A}$ is of Laplace type
with
\begin{equation}
(A_A)_B^{\ C} = 2 \omega_{;D}(\Sigma_A^{\ D})_B^{\ C},
\end{equation}
and
\begin{equation}
X_A^{\  B} = R_A^{\  B} - 2 \omega_{;A}^{\ \ B} + 
\omega_{;A}\omega^{;B}-\omega_{;C}\omega^{;C}\delta_A^B.
\end{equation}
We define the projection operators
\begin{equation}
P_- = \delta^A_{B} - N^A N_B, \quad P_+ = N^A N_B. \quad \chi = P_+-P_- =
\delta^A_B-2 N^A N_B
\end{equation}
and
\begin{equation}
{\cal S}= \left(-K - 2\omega_{;N}\right)P_+
\end{equation}
Tangential derivatives of the projection operators in this gauge are again zero
\begin{equation}
\chi_{|I} =0, \quad \chi_{|IJ} =0, \quad \chi_{|IJK} =0,
\end{equation}
as is the commutator of two covariant derivatives.
\begin{equation}
\Omega_{AB}=0.
\end{equation}
We can therefore consider the vector field as a collection of scalar fields,
with the four
tangential components having $X=\left(\omega''-2\omega'^2\right)e^{2 \omega}$
and obeying
Dirichlet boundary conditions. The
normal component has $X=\left(2\omega''-2\omega'^2\right)e^{2 \omega}$ and
obeys Robin boundary
conditions with ${\cal S}=-K-2 \omega_{;N}$.

The ghost field operator is given by the variation of the gauge fixing term
with respect to the
gauge parameter. This gives us $X=\left(\omega''-2\omega'^2\right)e^{2
\omega}$. BRST invariance
requires the ghost fields to obey Dirichlet boundary conditions. The ghost
fields
therefore cancel the contribution to the effective action from two of the
tangential components
of the vector field.

The coefficients $c_i$ and $d_i$ for the gauge field are shown in
Table~\ref{tab:coeff}. Here, the
effect of the ghost field has already been included.

\subsection{Fermion fields}

The squared Dirac operators $\Delta={\cal D}^* {\cal D}$ and
$\Delta^*={\cal D}{\cal D}^*$ are of Laplace type with
\begin{equation}
X=\frac14 R + m^2 \pm  \slash \partial m ,
\end{equation}
with the upper (lower) sign for $\Delta$ ($\Delta^*$). From now on, we
consider the operator $\Delta$ only, as the results for $\Delta^*$ are 
identical. 

We consider fermion mass terms of the form $m=b\omega' e^{\omega}$. The mass
term for the
fermion in the vector multiplet considered in Section~\ref{sec:fermion} is of
this 
form with $b=1/2$. 
The fermions obey
boundary conditions $\Pi_- \psi=0$, where 
\begin{equation} 
\Pi_\pm = \frac12(1\pm \Gamma^{11}).
\end{equation}
To write this in a covariant form, we should write the projection operators
$\Pi_\pm$ in terms
of $\Gamma^N$. However, since the outward unit normal is of opposite sign on
the two
boundaries, we must define the projection operators
\begin{equation}
P_\pm = \left\{\begin{array}{l}
               \frac12(1\mp  \Gamma^{N}) \quad \text{on} \ \ \partial {\cal M}^{(1)}\\
	       \frac12(1\pm  \Gamma^{N}) \quad \text{on} \ \ \partial {\cal M}^{(2)}
	       \end{array} \right.
\end{equation}	       
The boundary condition $P_+{\cal D}\psi=0$
can be written as $(\nabla_N - {\cal S})P_+\psi=0$, where	    
\begin{equation}
{\cal S} = (-\frac12 K - m)P_+ \quad \text{on} \ \ \partial {\cal M}^{(1)}, \quad
{\cal S} = (-\frac12 K + m)P_+ \quad \text{on} \ \ \partial {\cal M}^{(2)}.
\end{equation}
Tangential derivatives of the projection operators are non-zero, for example,
on $\partial {\cal M}^{(2)}$,
\begin{equation}
\chi_{|I}= \frac{1}{4} \Gamma_I K, \quad \chi_{|IJ} = -\frac{K^2}{16} \chi
\eta_{IJ}, 
\quad  \chi_{|IJK} = -\frac{K^3}{64} \eta_{IJ} \Gamma_K.
\end{equation} 
The commutator of two covariant derivatives is also non-zero,
\begin{equation}
\Omega_{AB}= i R_{AB}^{\ \ \ CD} \Sigma_{CD} \quad \Sigma_{CD}= -\frac i8
[\Gamma_C, \Gamma_D].
\end{equation}
After inserting these into the general form for the heat kernel coefficient, we
find
the results for the $c_i$ and $d_i$ listed in Table~\ref{tab:coeff}.

\begin{table}
\begin{ruledtabular}
\begin{tabular}{lccccc}
Field 		 & $c_1$ 	& $c_2$ 		& $c_3$ 		
& $c_4$ 		& $c_5$             \\\hline\hline

Dirichlet Scalar & $2880 \hz$ 	& $-8640\hz$		& $-8640\hz$ 		
& $138240\hz^2$ 	& $-103680\hz^2$    \\
                 &              &                       & $-46080\hz^2$        
&                 	&		    \\ \hline
		 
Robin Scalar     & $-2880 \hz$  & $8640\hz$ 		& $8640\hz$		
& $-138240\hz^2$  	& $103680\hz^2$     \\
                 &              & $+46080\hz\he$        & $+46080\hz^2$        
&  $-368640\hz\he^2$    & $+552960\he^2\hz$ \\
		 &		&			&			
&  $-138240\hz\he $	& $+368640\he^4$    \\ \hline
Vector           &  $-540$      & $540$                 &  $0$                 
&  $540$                & $-135$            \\ \hline
		 
Fermion   	 &    $+1440 b$	& $-2880 b^2$    		&   $0$			
& $-5760 b^3$		& $2880 b^4$   \\
&&&&&\\ 
&&&&&\\
\hline          

Field 		 & $d_1$ 	& $d_2$ 	& $d_3$ 	& $d_4$ 		
& $d_5$            \\ \hline\hline
Dirichlet Scalar & $-45$ 	& $150$		& $120$   	& $-\frac{135}{2}$ 	
& $-\frac{165}{8}$ \\
                 &              & $+3600\hz$   	& $+2880\hz$    & $-17280\hz$          
& $+3240\hz$        \\ \hline
		 
Robin Scalar     & $ 45$  	& $-120$ 	& $-120$	& $\frac{75}{2}$  	
& $\frac38$   	\\
                 &              & $-720\he$     & $-2880\hz$    &  $+20160\hz$    
& $-1080\hz$           \\
		 &              & $-5040\hz$    & 		&  $+1680\he $		
& $-120\he$		\\
		 &		&		&		& $+46080\he\hz$	
& $-69120\he\hz$	\\
		 &		&		&		& $+5760\he^2$		
& $-3840\he^2$		\\
		 &		&		&		& 			
& $-46080\he^3$	\\ \hline
Vector           &  $-45$       & $-675$        &  $-420$       & 
$\frac{2205}{2} $    & $-\frac{1107}{8}$            \\ \hline		 
Fermion   	 & $ 0 $ 	& $120$    	& $-1440 b$       & $-60$     		
& $-\frac{99}{2}$  \\
		 &		& $+2520b$	&		& $-480b$               
& $-60b$ \\
		 &		&		&		& $-2880b^2$ 		
& $-240b^2$ \\
		 &		&		&		&			
& $-1440b^3$
\end{tabular}
\end{ruledtabular}
\caption{Coefficients for cocycle function in five dimensions for flat
boundaries. The vector field result includes the
ghost fields, and the fermion result includes the fermion sign factor and the
factor of 4 for the number of degrees of freedom of the fermion. The scalar
mass squared is $\xi R$, the Robin boundary operator $(\partial_N+\eta K)$ and 
the fermion mass $m=b\omega'e^{\omega}$. The parameters $b$, $\hat\xi=\xi-\frac3{16}$
and  $\hat\eta=\eta-\frac38$ vanish for conformally covariant
operators.}\label{tab:coeff}
\end{table}

\bibliography{cocycle}

\end{document}